\documentclass[onecolumn]{aa} 
\usepackage{graphicx,color}
\usepackage{txfonts}
\usepackage{natbib}
\bibpunct{(}{)}{;}{a}{}{,} 

\def\aap{A\&A}

\def\apjl{ApJL}

\begin{document}
   \title{A disk inside the bipolar planetary nebula M2-9\thanks{Based on observations collected at the European Organisation for Astronomical Research in the Southern Hemisphere, Chile, ESO N: 079.D-146}}


   \author{F. Lykou\inst{1}
          \and
          O. Chesneau\inst{2}
          \and
          A.A. Zijlstra\inst{1}
                              \and
	A. Castro-Carrizo\inst{3}
          \and
          E. Lagadec\inst{4}
                    \and
          B. Balick\inst{5}
          \and
          N. Smith\inst{6}
          }


   \institute{Jodrell Bank Centre for Astrophysics, The Alan Turing Building, School of Physics \& Astronomy, University of Manchester, Oxford Rd, Manchester, M13 9PL, UK 
             \email{Foteini.Lykou@postgrad.manchester.ac.uk}
        \and
UMR 6525 H. Fizeau, Univ. Nice Sophia Antipolis, CNRS, Observatoire de la C\^{o}te d'Azur, 
Av. Copernic, F-06130 Grasse, France
\and
    Institut de RadioAstronomie Millim\'etrique (IRAM), 300 rue de la Piscine, F-38406 St. Martin d'Heres, France
   \and
    European Southern Observatory, Karl-Schwarzschild-Str. 2, D-85748 Garching bei M\"unchen, Germany
   \and
Astronomy Department, Box 351580, University of Washington, Seattle WA 98195, U.S.A.
\and
 Steward Observatory, University of Arizona, 933 North Cherry Avenue, Tucson, AZ 85721, U.S.A.
              }

   \date{Received December 10, 2009; accepted Month? ??, 2010}

 
  \abstract
  {}
           {Bipolarity in proto-planetary and planetary nebulae is associated with events occurring in or around their cores. Past infrared observations have revealed the presence of dusty structures around the cores, many in the form of disks. Characterising those dusty disks provides invaluable constraints on the physical processes that govern the final mass expulsion of intermediate-mass stars. We focus this study on the famous M2-9 bipolar nebula, where the moving lighthouse beam pattern indicates the presence of a wide binary. The compact and dense dusty core in the center of the nebula can be studied by means of optical interferometry. }
           {M2-9 was observed with VLTI/MIDI at 39-47~m baselines with the UT2-UT3 and UT3-UT4 baseline configurations. These observations are interpreted using a dust radiative transfer Monte Carlo code.}
           {A disk-like structure is detected perpendicular to the lobes and a good fit is found with a stratified disk model composed of amorphous silicates. The disk is compact, 25$\times$35~mas at 8$\rm \mu m$, and 37$\times$46 mas at 13$\rm \mu m$.  For the adopted distance of 1.2~kpc, the inner rim of the disk is $\sim$15~AU. The mass represents a few percent of the mass found in the lobes. The compactness of the disk puts strong constraints on the binary content of the system, given an estimated orbital period 90-120yr. We derive masses of the binary components between 0.6--1.0M$_{\sun}$ for a white dwarf and 0.6--1.4M$_{\sun}$ for an evolved star. We present different scenarios on the geometric structure of the disk accounting for the interactions of the binary system, which includes an accretion disk as well. }
           {}

   \keywords{planetary nebulae --
                post-AGB stars --
                infrared interferometry --
                disks
               }

   \maketitle
%

\section{Introduction}
Intermediate mass stars (1--8M$_{\sun}$) undergo extreme mass-loss during the late stages of their evolution. The mechanisms that dominate those stellar outflows are also responsible for the shaping of the ejecta. Hints of asymmetries are numerous in many stars characterizing these late stages. Almost all of planetary and proto-planetary nebulae~\citep{2007AJ....134.2200S} observed by the Hubble Space Telescope\footnote{http://www.astro.washington.edu/users/balick/WFPC2/index.html} exhibit asymmetries in their shape, while bipolarity and equatorial dusty disks are very common as well.

Most proposed shaping mechanisms for bipolar nebulae involve binarity~\citep{2007BaltA..16...79Z}, since a source of angular momentum is required to drive the ejecta in directions perpendicular to the orbital plane. Interaction between the ejecta and the orbiting companion can lead to the formation of torii, disks and jets, the latter perpendicular to the binary system's plane. During common envelope evolution, accretion disks can be formed leading to the ejection of jets, thus forming bipolar nebulae~\citep{1994ApJ...421..219S}. Disks can be formed during wind accretion preceding common envelope stage as well~\citep{1998ApJ...497..303M}, but are too low mass to be the direct cause of the shaping of the older nebula~\citep{2007A&A...473L..29C}. The disks appear to be a remnant of the shaping process. 

Disks are rather small in size (a few milliarcseconds), compared to the surrounding nebulae (many arcsecs), and their dust re-emits stellar radiation in the infrared. Observations of such small structures near evolved stars can only be performed with the use of high-angular resolution techniques like the infrared Very Large Telescope Interferometer (VLTI), i.e.~planetary nebulae:~\citet{2006A&A...455.1009C, 2007A&A...473L..29C, 2006A&A...448..203L}; VLTP star:~\citet{2009A&A...493L..17C}; Mira stars:~\citet{2006ApJ...652..666T, 2007ApJ...670L..21T, 2008ApJ...689.1289T}; Carbon stars:~\citet{2008A&A...478..809O, 2008A&A...490..173O, 2008A&A...482..561S, 2007A&A...467.1093D}; OH/IR stars:~\citet{2007A&A...474L..45D, 2007A&A...467.1093D, 2005A&A...435..563C}; post-AGB stars:~\citet{2006A&A...450..181D, 2006ApJ...646L.123M}; proto-planetary nebulae:~\citet{2008A&A...489..195M, 2006A&A...445.1015O}.
\subsection{Twin-jet nebula M2-9}

Two of the best-studied examples of elongated and axisymmetric bipolar planetary nebulae are M2-9 and Menzel 3. They present tightly pinched waists and symmetric lobes. They are also spectroscopic twins at visual and IR wavelengths~\citep{2003MNRAS.342..383S, 2005AJ....129..969S}. This paper is focused on M2-9. The lobes of the nebula extend up to 22\arcsec~at each side of the bright central core.~\citet{1997A&A...319..267S} found that the nebula has polar knots extended up to 52\arcsec~from the central source (core). The core is dominated by H$\rm \alpha$ emission scattered by dust in the visual, as well as free-free emission in radio (diameter $\lesssim3$\arcsec). 

A lighthouse beam is seen inside the bipolar lobes and a series of observations at different epochs revealed that it is a by-product of the mechanism responsible for the shaping of the planetary nebula. The lighthouse effect was witnessed by knots in the nebula known as early as 1952~\citep{1972ApJ...174..583A}, was monitored from 1989 to 2007, revealing a period of $\sim$100 years. This feature can be created by orbital motion and is the best evidence that the core engulfs a binary system. In addition to that, the high-excitation lines seen in M2-9 suggest that the illuminating source is a hot and compact star, i.e. a white dwarf (WD)~\citep{1997A&A...319..267S, 2000ASPC..199..259L, 2001ApJ...552..685L}, {\it not} to be confused with the primary star whose ejecta now constitute the nebula. In such a scenario, dust originating from the primary will settle in the binary system's orbital plane in the form of a disk, torus or spiral.


The distance of M2-9 is uncertain. Many authors quote a distance of 1,000~pc in the absence of a tight constraint~\citep{1994ApJ...437..281H, 2001ApJ...552..685L, 2005AJ....129..969S}.~\citet{1997A&A...319..267S} propose a much closer distance of 640$\pm$100pc,  based on expansion parallax measurements of the outer lobes. We derive a larger distance in this paper, confirmed by R. Corradi (private communication) who reanalyzed the data of~\citet{1997A&A...319..267S}. For all the mass estimates described below, the assumed distance\footnote{All mass estimates with an assumed distance of 1kpc, except for~\citet{1989ApJ...345..306L} (2.37kpc)} was different from the one presented in this paper.

\citet{1990A&A...227..188B} mentioned a total amount of molecular gas in the nebula higher than $10^{-3}$~M$_{\sun}$.~\citet{1984ApJ...287..353F} gave a gaseous mass of 0.21~M$_{\sun}$~and mass of dust $2.1\times10^{-3}$~M$_{\sun}$ for the entire nebula, but~\citet{1989ApJ...345..306L} estimated a higher total dust mass by IRAS observations ($\sim5.4\times10^{-3}$~M$_{\sun}$).

The core of M2-9 was resolved using millimetric interferometry (Plateau de Bure, 5\arcsec$\times$3\arcsec~beam), showing that the molecular gas is distributed in a ring-like structure extended about 6\arcsec~E-W. A kinematical age $\sim$2,100 years was derived from the expansion of the CO gas in the nebula~\citep{1997A&A...324..624Z}.

Observations by~\citet{2005AJ....129..969S} within the last decade showed that in the core the mass of warm gas is $9.2\times10^{-4}$~M$_{\sun}$~and that of warm dust $4\times10^{-6}$~M$_{\sun}$, while in the lobes the mass of cold gas is 0.78~M$_{\sun}$ and that of cool dust is $3.4\times10^{-3}$~M$_{\sun}$.

We present observations of the core of M2-9 in the near- and mid-infrared (Section 2). We have detected a dusty disk in the core with the use of high-angular resolution techniques. The geometric constraints of the disk were defined with the use of radiative transfer models. Our results are presented in Sections 3 \& 4. A detailed discussion can be found in Sections 5-6, where we elaborate scenarios on the disk's structure. In Section 7, we compare our findings with the disk of Menzel 3. Our conclusions are in the final section.


\section{Observations}

\subsection{VLTI}
M2-9 was observed in the N band (7.5-13.5$\rm \mu m$) with MIDI~\citep{2003Ap&SS.286...73L, 2004A&A...423..537L}, the mid-infrared recombiner of the VLTI with the use of only two Unit Telescopes (UTs). We used a typical MIDI observing sequence as described in~\citet{2003Ap&SS.286...85P}. MIDI can make single-dish acquisition images with a field-of-view of about 3$\arcsec$ with a spatial resolution of about 0.25$\arcsec$ at 8.7$\rm \mu m$, and provides a flux calibrated spectrum at low spectral resolution (in this case R=30) and several visibility spectra from the sources~\citep{2005A&A...435.1043C, 2005A&A...435..563C}, i.e. spectrally dispersed information on the spatial extension of the source. Only four visibility measurements could be obtained, due to bad weather conditions.

The observations of M2-9 with MIDI were performed in April, June and August 2007 and March 2008 in the so-called SCI\_PHOT mode, meaning that the photometry of each telescope is recorded simultaneously to the fringes. The errors, including the internal ones and the ones from the calibrator diameter uncertainty, range from 5\% to 10\%. The accuracy of the absolute flux calibration is about 10-13\%. For the flux calibration, HD 163917 (F$_{12\rm \mu m}=16.54$~Jy) was used. A Cohen template for a K0III star was used (HD 180711, F$_{12\rm \mu m}=21.35$~Jy). The log of the observation is given in Table~\ref{table-log}. We used two different MIDI data reduction packages: {\tt MIA} developed at the Max-Planck-Institut f\"ur Astronomie and {\tt EWS} developed at the Leiden Observatory\footnote{Available at http://www.strw.leidenuniv.nl/~nevec/MIDI/index.html}.

\begin{table}
\centering
        \begin{caption}{Observing log}\label{table-log}\end{caption}
        \begin{tabular}{llccc}\hline\hline
        OB   & Time  & Base & \multicolumn{2}{c}{Projected baseline}  \\
         & & & Length  & PA   \\
         &&&[m] & [degrees] \\
        \hline
        M2-9\_1  & 2007-04-11 & U2 - U3 & 40.2 & 107.3  \\
        M2-9\_2  & 2007-06-28 & U2 - U3 & 47.3 & 126.0 \\
        M2-9\_3  & 2007-08-28 & U2 - U3 & 39.7 & 44.0  \\
        M2-9\_4  & 2008-03-28 & U3 - U4 & 45.1 & 39.0  \\
        \hline
        \end{tabular}
\\
        {Calibrators: HD 116870 K5III $2.44\pm0.12$mas, HD 163917 G9III 2.75$\pm$0.11mas, HD 167618 M3.5III 11.33$\pm$0.1mas, HD175775 G9III 3.26$\pm$0.23mas, HD 152820 K5III 2.57$\pm$0.34mas
        }
\end{table}

\begin{figure}[!h]
  \centering
    \includegraphics[width=0.9\textwidth]{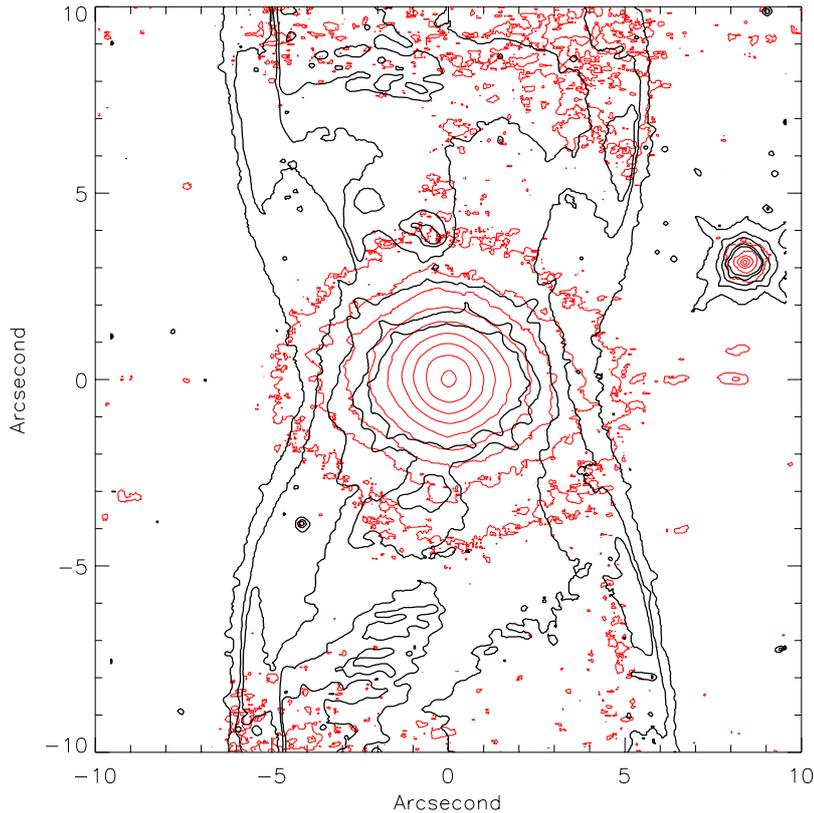}
\vspace{-10cm}
    \caption[]{Composition of HST/STIS (black) and NACO/Ks (red) images. North is up and East is on the left. The infrared emission contours follow the lines of the HST scattering in the visible, not only in the core but in a N-W region as well. The unresolved core is less than 0.1\arcsec~in Ks. Contours from the center to low levels are 95, 10, 2, 0.8, 0.4, 0.2, 0.1, 0.05, 0.025 and 0.015\% of the maximum, respectively. Noise appears in the last two lower levels and those reveal the extended nebula as it is evident on the northern and southern part of the image. The diffuse light is increasing smoothly from the 0.05 to the 10\% level.
        \label{fig:hst-naco} }
\end{figure}

\subsection{NACO}

We observed M2-9 with the near-infrared adaptive optics instrument NACO~\citep{2003SPIE.4841..944L,2003SPIE.4839..140R} as well, using three broad-band filters centred respectively at 2.18$\rm \mu m$ (Ks), 3.8$\rm \mu m$ (L$'$) and 4.78$\rm \mu m$ (M$'$). The Ks data were taken during the night of 3 July 2006, and the L$'$-band and M$'$-band data on 18 August 2006. These observations were complemented with close-by PSF calibrations using HD155078 (F5V) in L$'$ and HD156971 (F1III) in M$'$. We used the S27 (Ks) and L27 (L$'$, M$'$) camera mode to obtain a field of view of 28\arcsec$\times$28\arcsec and the pixel scale of 27.1~mas per pixel. Jittering was used to remove the sky and for the M$'$ observations chopping was also used. The reduction procedure followed for MIDI and NACO data of M2-9, is the same as the one described for Mz3 in~\citet{2007A&A...473L..29C}.

Our Ks, L$'$ and M$'$ observations of Mz3~\citep{2007A&A...473L..29C} and M2-9 with NACO revealed an unresolved core at diameters less than 0.1\arcsec. Figure~\ref{fig:hst-naco} is a composite image of a Hubble Space Telescope (HST) STIS/CCD image in the visible and our Ks-band image, the latter being an improved near-infrared view of the source compared to~\citet{1988A&A...196..227A}. One can notice a broad near-infrared emission around the core of M2-9 (diameter$<5$\arcsec) and some enhanced emission in the lobes as well ($\sim$10\arcsec~from the core; more prominent at N-W). These are attributed to scattering. Near-IR photometry gave m$_{L'}$=4.12 and m$_{M'}$=2.5 and these are included in Fig.~\ref{fig:SED}.

\subsection{ISO}

We have reduced and analysed archived\footnote{http://iso.esac.eso.esa.int/ida/} data from ISOCAM~\citep{1996A&A...315L..32C}, the Infrared Space Observatory's (ISO) infrared camera. The satellite was mounted with a 60cm Ritchey-Chr\'etien Cassegrain telescope. Ten exposures have been recorded in 1996 for a range of wavelengths from 3.0 to 14.9$\rm \mu m$ with both short-wave (SW) and long-wave (LW) filters (bandwidths stated in Table~\ref{tab:isocam}). Each frame is 32$\times$32 pixels with an effective field-of-view 45\arcsec$\times$45\arcsec~and the spatial scale of each pixel is 1.5\arcsec. The standard software {\tt CIA} was used to reduce and analyse the data (Table~\ref{tab:isocam}) in the procedure described in the manual. Photometric errors were consistent for each detector. In ISOCAM the physical detector pixel size is 1\arcsec\,and the Airy disk pattern increases with wavelength from 1\arcsec~to 3.5\arcsec~\citep[see][p. 44]{2003ESASP1262.....B} thus the core of M2-9 was not resolved by this instrument. The broad infrared emission from the core did not permit the detection of any other significant structure in its vicinity either.

\begin{table}
\centering
        \begin{caption}{ISOCAM photometry}\label{tab:isocam}\end{caption}
        \begin{tabular}{ccccc}\hline\hline
        $\rm\lambda_{ref}$ [$\rm\mu$m] & BW [$\rm\mu$m] & detector & Flux [Jy] & $\rm \sigma_{flux}$~[mJy] \\
        \hline
        3.0	&1.00 & SW3 & 4.6	& 9.5	\\
        3.7	&0.55 & SW6 & 7.9	& 9.6	\\
        4.5	&0.30 & SW7 & 13.3	& 9.5	\\ \hline
        4.5	&1.00 & LW1 & 16.1	& 16.7	\\
        6.0	&1.00 & LW4 & 27.6	& 16.7	\\
        6.8	&0.50 & LW5 & 32.1	& 16.4	\\
        7.7	&1.50 & LW6 & 37.2	& 16.6	\\
        9.6	&2.20 & LW7 & 34.7	& 16.7	\\
        11.3	&1.30 & LW8 & 42.3	& 16.7	\\
        14.9	&2.00 & LW9 & 50.6	& 16.6	\\
        \hline
        \end{tabular}
\end{table}

An infrared spectrum has also been retrieved from ISO Short Wavelength Spectrometer (SWS) using short- and long-wavelength gratings~\citep{1996A&A...315L..27K}. The aperture area varies with wavelength range and grating: it is 14\arcsec$\times$20\arcsec~for 2.38--12.0$\rm \mu m$, 14\arcsec$\times$27\arcsec~for 12.0--27.5$\rm \mu m$, 20\arcsec$\times$27\arcsec~for 27.5--29.0$\rm \mu m$ and 20\arcsec$\times$33\arcsec~for 29.0--45.2$\rm \mu m$ (see Fig.~\ref{fig:SED}).

\subsection{Spitzer}

\begin{figure}[!h]
	\centering
	\includegraphics[width=0.7\textwidth,height=10.5cm]{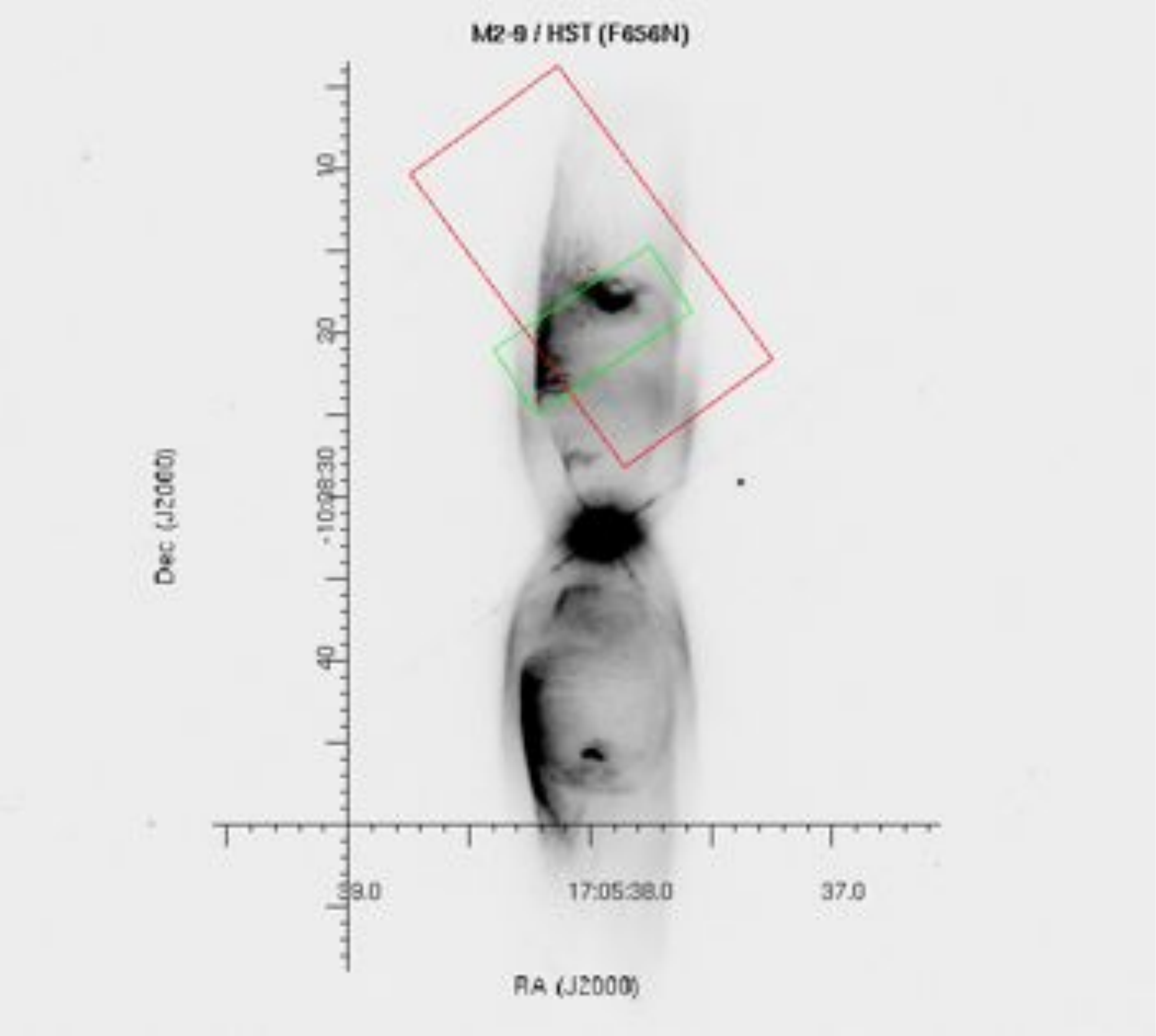}
	\caption[]{Positioning of the Spitzer/IRS slits over a F656N exposure of M2-9 (Hubble Space Telescope Legacy Archive). Both slits were centered on top of the northern lobe, the Short-High (green; 4.7\arcsec$\times$11.3\arcsec) at a PA=131.2\degr and the Long-High (red; 11.1\arcsec$\times$22.3\arcsec) at a PA=46.1\degr. North is up, East is left. The underlying H$\rm\alpha$ image comes from 1996.  By the time that Spitzer observed M2-9, the underlying structure of M2-9 was rather different, and the nebular structure inside the green aperture had changed.  \label{fig:slits} }
\end{figure}

Calibrated data were extracted from the Spitzer Space Telescope\footnote{http://ssc.spitzer.caltech.edu/} Archive. M2-9 has been observed with both the InfraRed Array Camera (IRAC) and the InfraRed Spectrograph (IRS). Unfortunately, the core's strong infrared emission has saturated IRAC rendering those observations useless. \par
\begin{figure}[!h]
\centering
	\includegraphics[width=\textwidth]{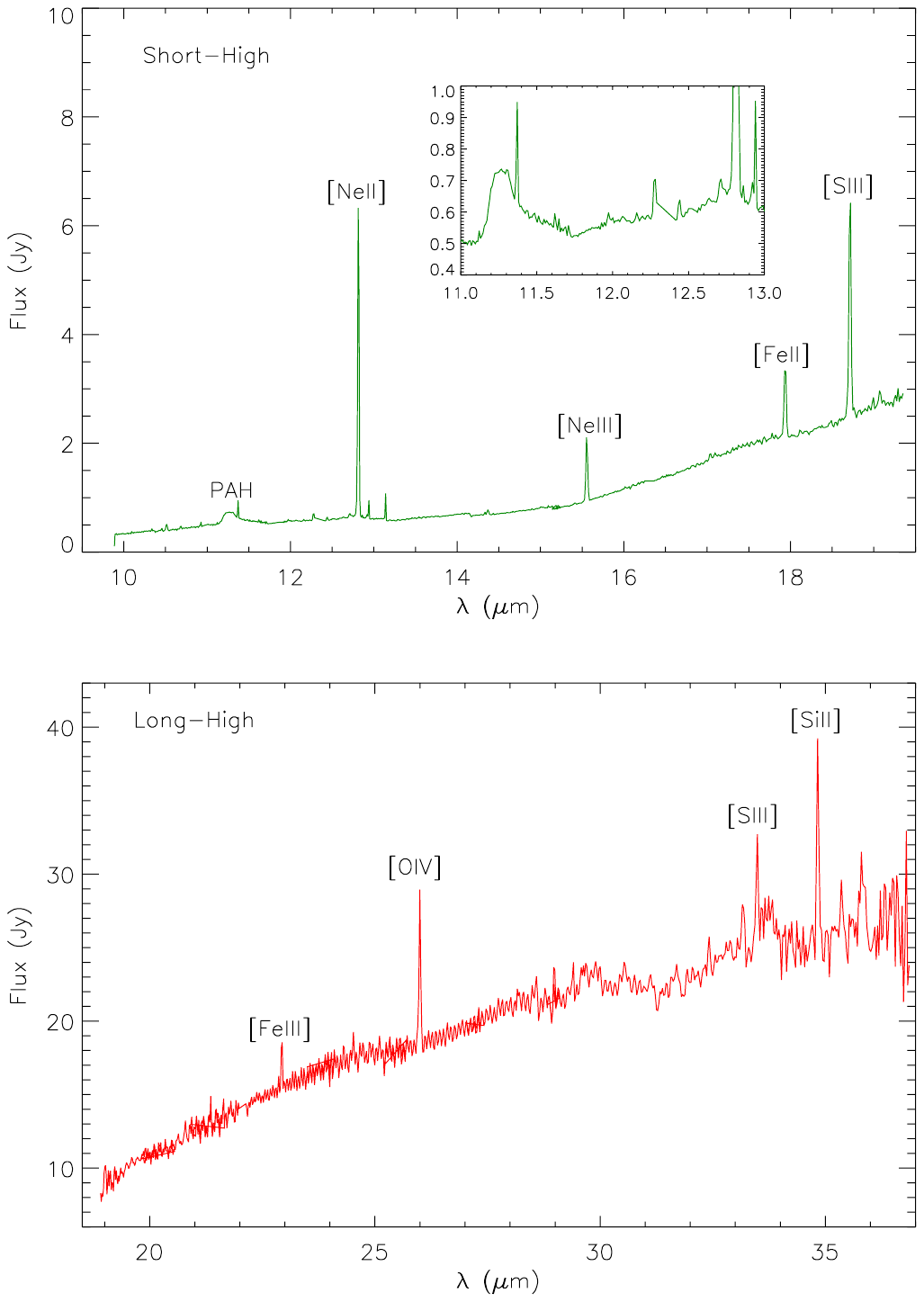}
	\caption[]{High-resolution spectra of the northern lobe of M2-9 by IRS/SPITZER. Upper curve corresponds to the Short-High slit and lower curve to the Long-High (for slit orientation, see Fig.~\ref{fig:slits}). The inset in the upper plot is a close-up of the PAH emission features. \label{fig:spitzer} }
\end{figure}

High-resolution spectra (Fig.~\ref{fig:spitzer}) by IRS revealed important information on the nebular dusty and gaseous chemistry. Two slits were positioned over the northern lobe and avoided the core (Fig.~\ref{fig:slits}): Short-High slit (9.9--19.6$\rm \mu m$) is 4.7\arcsec$\times$11.3\arcsec~at PA=131.2\degr~and Long-High slit (18.7--37.2$\rm \mu m$) is 11.1\arcsec$\times$22.3\arcsec~at PA=46.1\degr.

\subsection{AKARI}

We include mid-infrared data from the AKARI satellite\footnote{http://www.ir.isas.jaxa.jp/AKARI/} IRC Point Source Catalogue archive~\citep{2007PASJ...59S.369M}. M2-9 has been observed with the InfraRed Camera~\citep{2007PASJ...59S.401O} at two infrared bands, namely S9W (6.7--11.6$\rm\mu m$) and L18W (13.9--25.6$\rm\mu m$) with respective effective wavelengths at 9 and 18 $\rm \mu m$ (see Fig.~\ref{fig:SED}).


\section{Results}

\subsection{Spectral Energy Distribution}

The integrated flux, as detected by MIDI (field-of-view: $\sim$0.4\arcsec~with UTs), in Fig.~\ref{fig:SED} clearly shows a 20\% difference between the MIDI and the ISO spectra. Since the rectangular aperture of ISO (14\arcsec$\times$22\arcsec) was positioned over the central area of M2-9 (excluding the largest part of the lobes) and it increased in size for bands larger than 12$\rm \mu m$ (including part of the lobes near the central source), it is expected that ISO detected a more extended source than MIDI. The fact that both spectra have similar shape and only slightly different flux levels, indicates that they both detected almost the same compact source.

Both ISO and MIDI spectra, shown in Fig.~\ref{fig:SED}, have been dereddened using the~\citet{1979ARA&A..17...73S} law for $A_{V}=2.5$~\citep{1998RMxAC...7..171T}.

The two Spitzer slits were positioned on top of the northern lobe, but they have probed different parts of the nebula. In more detail (Fig.~\ref{fig:spitzer}):
\begin{itemize}
	\item {\it Short-High}:\\
The spectrum is very weak compared to the continuum of both MIDI and ISO, which were positioned over the core, but there are many interesting features. There is a broad PAH emission at 11$\rm \mu m$ and much fainter broad regions at 12.3 and 12.8$\rm \mu m$.~[NeII], [NeIII], [FeII] and [SIII] lines are seen at 12.8, 15.6, 17.9 and 18.7$\rm \mu m$, respectively. No other PAHs have been detected near the core by ISO, whereas it is unclear whether the 11.4$\rm \mu m$ emission at 10\% of the continuum of MIDI spectrum originates from PAH (Fig.~\ref{fig:forst}). This suggests that the 11$\rm \mu m$ feature resides in the lobe.\\
	\item {\it Long-High}:\\
There is an indication of a broad feature at 33.5$\rm \mu m$ (forsterite) but the neighboring broad shallow absorption from 30-34$\rm \mu m$ suggests that they may both not be real. Prominent gaseous emission lines are also present: [FeIII]~(22.9$\rm \mu m$), [OIV]~(25.9$\rm \mu m$), [SIII]~(33.48$\rm \mu m$) and [SiII]~(34.8$\rm \mu m$). All those require a high excitation state and would be expected near a white dwarf.
\end{itemize}


\subsection{Interferometry}

\begin{figure}[!h]
  \centering
   \includegraphics[width=0.75\textwidth, angle=90]{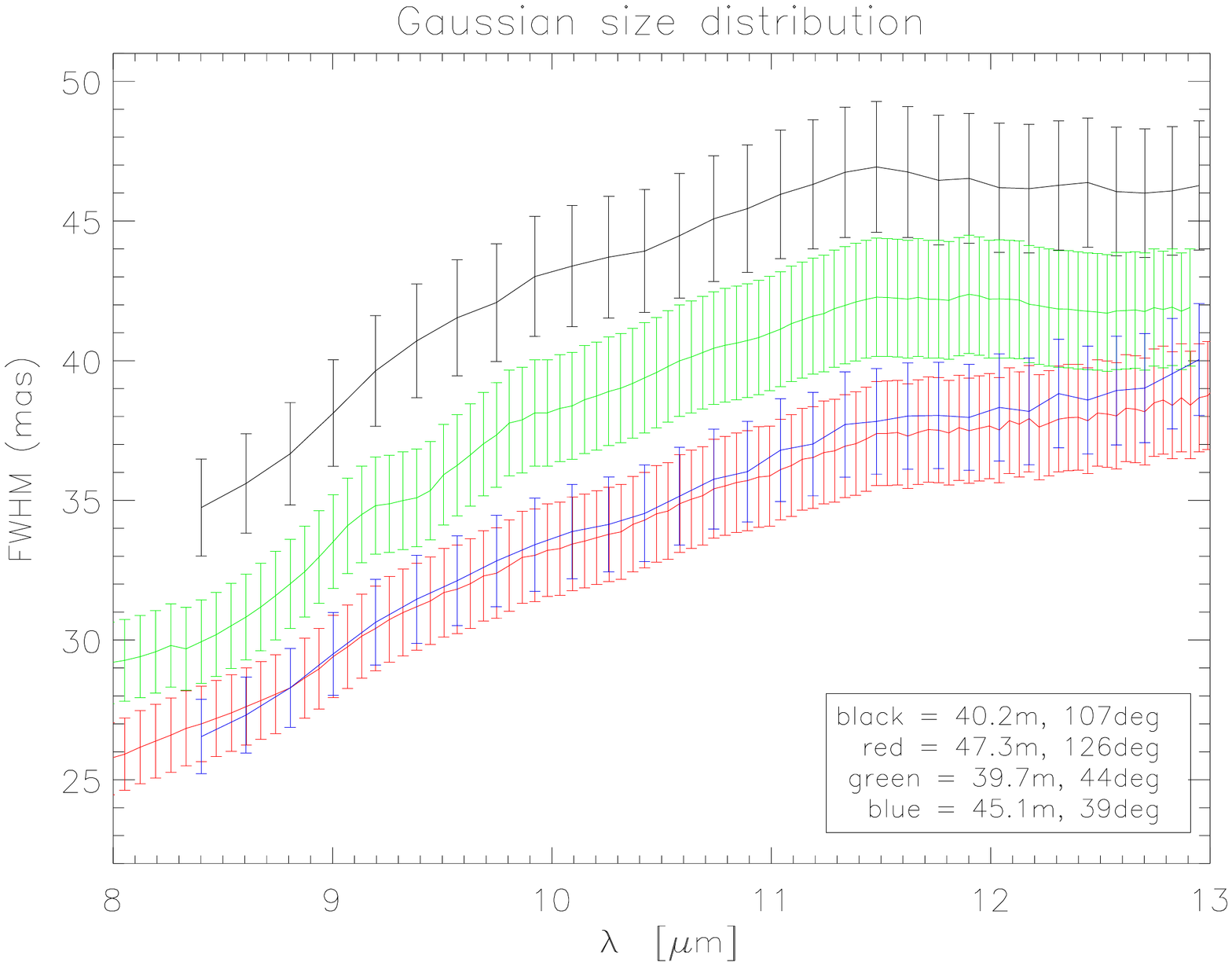}
    \caption[]{Gaussian size distribution of the disk ($\pm$5\%) for each baseline (Table~\ref{table-log}) in N band implementing the dusty structure's evolution in the mid-infrared. 
        \label{fig:gauss}}
\end{figure}

Our VLTI observations of the core of M2-9 reveal the presence of a dusty flat-like structure. Visibilities decrease as the P.A. of the projected baselines increases (Fig.~\ref{fig:gauss}), suggesting an increase of the dust opacity. Thus, MIDI has probed the existence of a flattened structure along the equatorial plane of the nebula.

The core is significantly resolved by the 40-50m baselines, with visibilities of the order of 0.1-0.2. Our observations showed that the flattened structure is compact, typically 25$\times$35~mas at 8$\rm \mu m$, and 37$\times$46 mas at 13$\rm \mu m$ (Fig.~\ref{fig:gauss}). The dusty structure is more elongated along the planetary nebula's equatorial plane, compared to the direction of the lobes. Therefore, the disk's geometrical size can be constrained (see Sections 4 \& 6). 

\begin{figure}[!h]
 \centering
   \includegraphics[width=\textwidth]{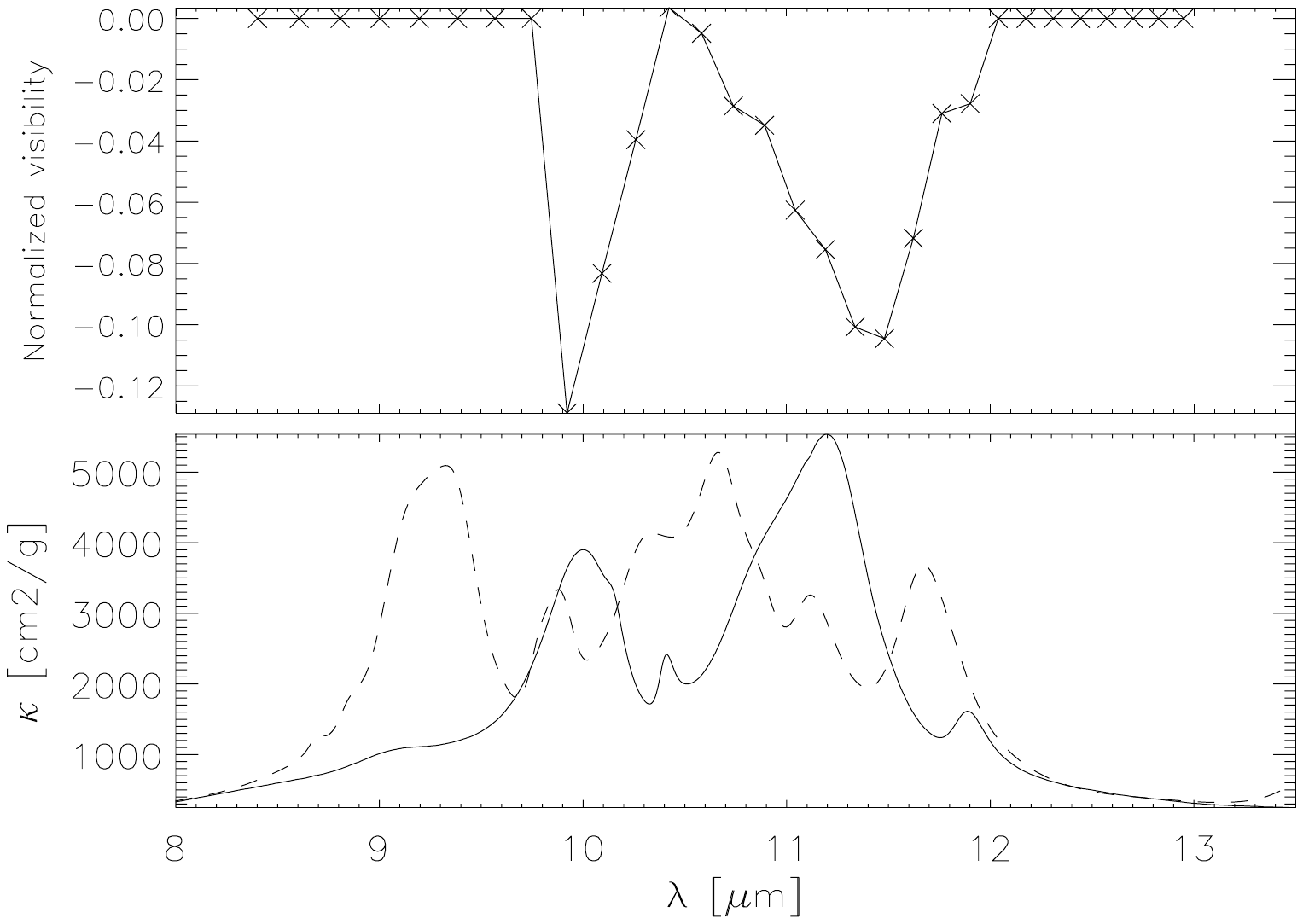}\vspace{-8cm}
    \caption[]{{\it Top}: Normalized visibility vs. wavelength for M2-9 revealing two crystalline silicate absorption bands at 9.8 and 11.4$\rm \mu m$; they are identified as forsterite. For the sake of clarity only the nearest to the equatorial plane measurement (M2-9\_1: 40.2m, 107.3\degr) is shown here. {\it Bottom}: Mass absorption coefficients for forsterite (solid) and enstatite (dash) taken by~\citet{2000A&A...363.1115K, 2006A&A...449..583K}
        \label{fig:forst}}
\end{figure}

Subtracting the continuum from each spectrally dispersed visibility, one may enhance subtle visibility fluctuations that could not be distinguished otherwise. Significant signatures of forsterite (Mg$_{2}$SiO$_{4}$) are observed as small ($\sim5-10\%$) dips in the visibility curves (Fig.~\ref{fig:forst}) at 9.8 and 11.4$\rm \mu m$~\citep{2002A&A...382..222M}. We confirmed the validity of this, since both features were present in the uncalibrated visibilities and they were absent in the calibrator data. In addition, the peaks nearly match the emission curves of forsterite grains measured in the lab by~\citet{2000A&A...363.1115K, 2006A&A...449..583K}. The visibility decrease at typically 10\% shows that the forming regions of these features are resolved and thus slightly more extended than the continuum and amorphous silicate regions (as already observed in Young Stellar Objects,~\citet{2004Natur.432..479V}).

\section{Physical parameters of the disk}

In this section we present the modelled disk and the constraints introduced by the fitting procedure.
\subsection{Model}

The 3D radiative transfer code MC3D, which is based on the Monte Carlo method and solves the radiative transfer problem self-consistently, has been used in this work~\citep{1999A&A...349..839W, 2003CoPhC.150...99W}. It simulates the temperatures produced by dusty configurations and creates as observables spectral energy distributions and wavelength-dependent images of the dusty environments, as well as polarisation maps. The density law used in this model is
\begin{equation}
	\rho\left(r,z\right)=\rho_{0}\left(\frac{R_{*}}{r}\right)^\alpha \exp{\left[-\frac{1}{2}\left(\frac{z}{h\left(r\right)}\right)^2\right]}
\end{equation}
\begin{equation}
	h\left(r\right)=h_{0}\left(\frac{r}{R_{*}}\right)^\beta
\end{equation}
\noindent
where {\it r} is the radial distance in the disk midplane, $R_{*}$ is the stellar radius, $\beta$ is the vertical density parameter, $\alpha$ is the density parameter in the midplane and $h_{0}$ is the scale height at a given distance from the star~\citep{1973A&A....24..337S,2002ApJ...564..887W} . We have used the distribution of \cite{1977ApJ...217..425M} for the grain sizes: $n(b)\propto b^{-3.5}$, where {\it b} is the grain size, assuming that the dust grains have spherical symmetry. 

The dust residing around evolved stars may be oxygen- or carbon-rich (O-, C-rich), or even a combination of both. This is determined by the abundance of each element during the AGB phase. Spectroscopic observations did not reveal any $3.3\rm \mu m$ poly-aromatic hydrocarbons (PAHs, i.e. signature of C-rich chemistry) in the core of M2-9~\citep{2008ApJ...676..408S}, but other PAH features are present in the lobes (this work, Section 3.1). The nebula is O-rich with $C/O<0.5$~\citep{2001MNRAS.323..343L}, thus the dust is expected to be O-rich. The model contained only amorphous silicate dust, since the exact percentage of crystalline silicates that is locked within the disk is unknown.

We have chosen a certain position angle (P.A.) of the gaseous nebula to fit the observed visibilities. According to~\citet{1999AJ....118.2919P}, the P.A. of M2-9 is -2\degr. We support this estimation from the positions of the two ansae at $\sim$52\arcsec~from the core and finding the best fit at the same position angle (Table~\ref{tab:param}).

\subsection{Fitting results}

Following the success of fitting a disk model for the core of Mz3 and knowing that those two objects are similar, we chose parameter ranges for the disk in M2-9 similar to those of Mz3 (Table~\ref{tab:param}). A more in-depth presentation on general MC3D fitting, can be found in~\citet{2009A&A...505.1167S}.

\begin{figure*}[!h]
 \centering
    \includegraphics[width=\textwidth]{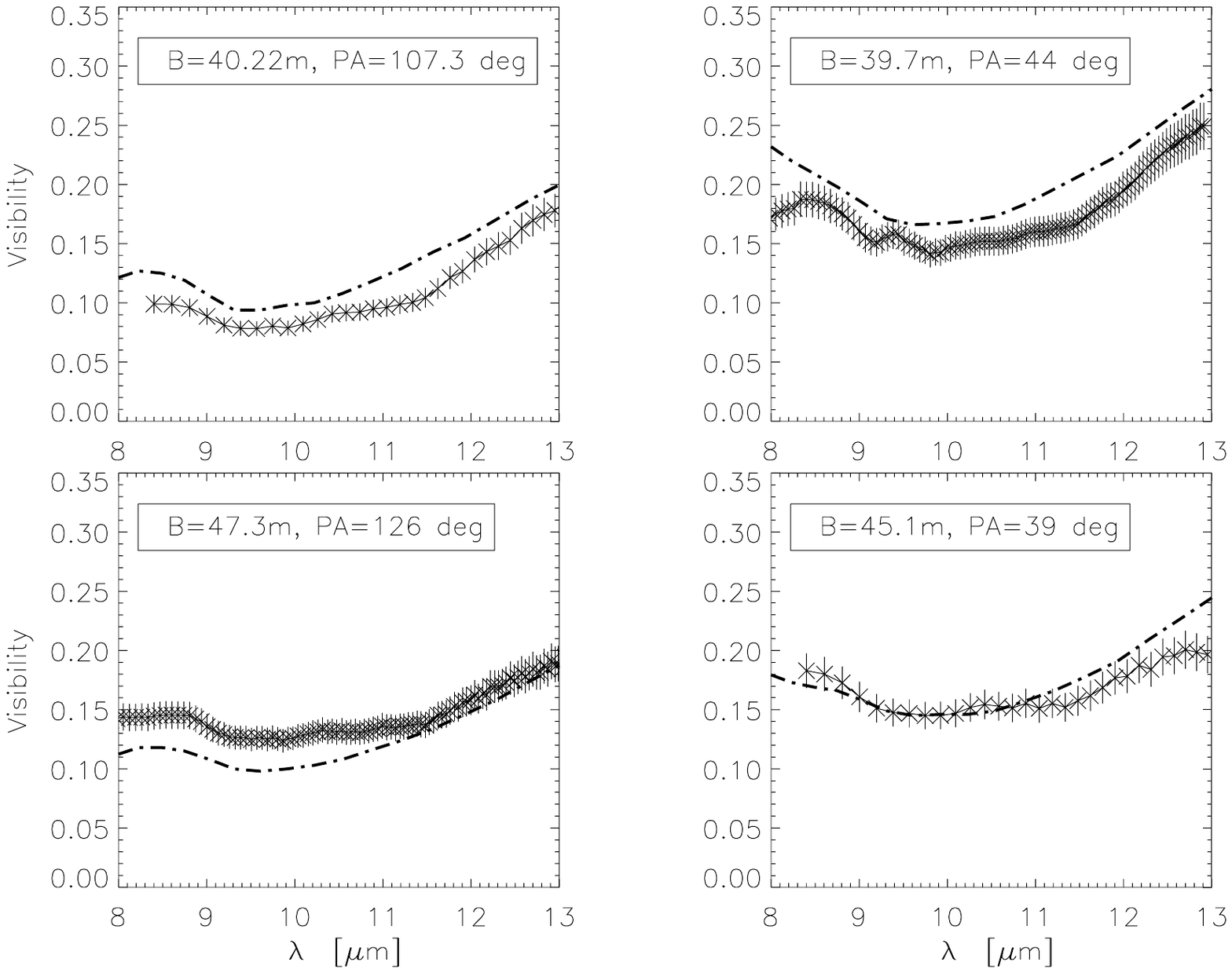}
	\vspace{-7cm}
 \caption[]{MIDI visibilities, compared with the best model (dashed-dotted lines, $\rm \chi^{2}$=3.17, see Table~\ref{tab:param}). The visibility curves are for the baselines M2\_9-1 to M2\_9-4 from top to bottom and left to right (Table~\ref{table-log}). Individual $\rm \chi^{2}$ are 2.61, 5.33, 3.05 and 1.69 for each curve respectively. \label{fig:VIS}}
\end{figure*}

\begin{itemize}
	\item The disk's {\it inner rim} radius was chosen by fitting the observed visibilities with Gaussian uniform disks, thus estimating their diameters in milliarcseconds, and converting those into physical units.
	\item We adopted an initial value for the {\it mass of warm dust} derived by \citet{2005AJ....129..969S}, but it was increased in order to recreate the silicate absorption features. Parameter $\alpha$ was fixed likewise to a certain value, where we would see silicates in absorption and at flux levels between those set by ISO and MIDI spectra.  In this paper, we do not probe the large amount of cold dust that was observed by~\citet{2005AJ....129..969S}.
	\item Low visibilities indicated a resolved structure, or else a very dense medium, so a {\it scale height, h$_{0}$} higher than 20~AU, was chosen initially, with increments of 5AU, and a moderate {\it flaring}, $\beta$, with steps of 0.01, in order to fit the observed visibilities (Fig.~\ref{fig:VIS}).
\end{itemize}

With the use of MC3D, the reconstruction of the disk's image was made. The models of its spectral energy distribution (SED) and visibility amplitudes have been compared to the MIDI data and the observed SED. After careful examination of all possible parameters, the model with the best fit, is one that resembles the best geometrical shape for a flared disk. The model's parameters are shown in Table~\ref{tab:param}.

We were able to define a lower and an upper limit to constrain the size of the disk's structure from a theoretical basis and from our observational data. A projected baseline on the equatorial direction of M2-9 would probe the full size of the disk. In our case, this telescope configuration was not accomplished.

\begin{figure*}[!ht]
\centering
	\includegraphics[width=\textwidth]{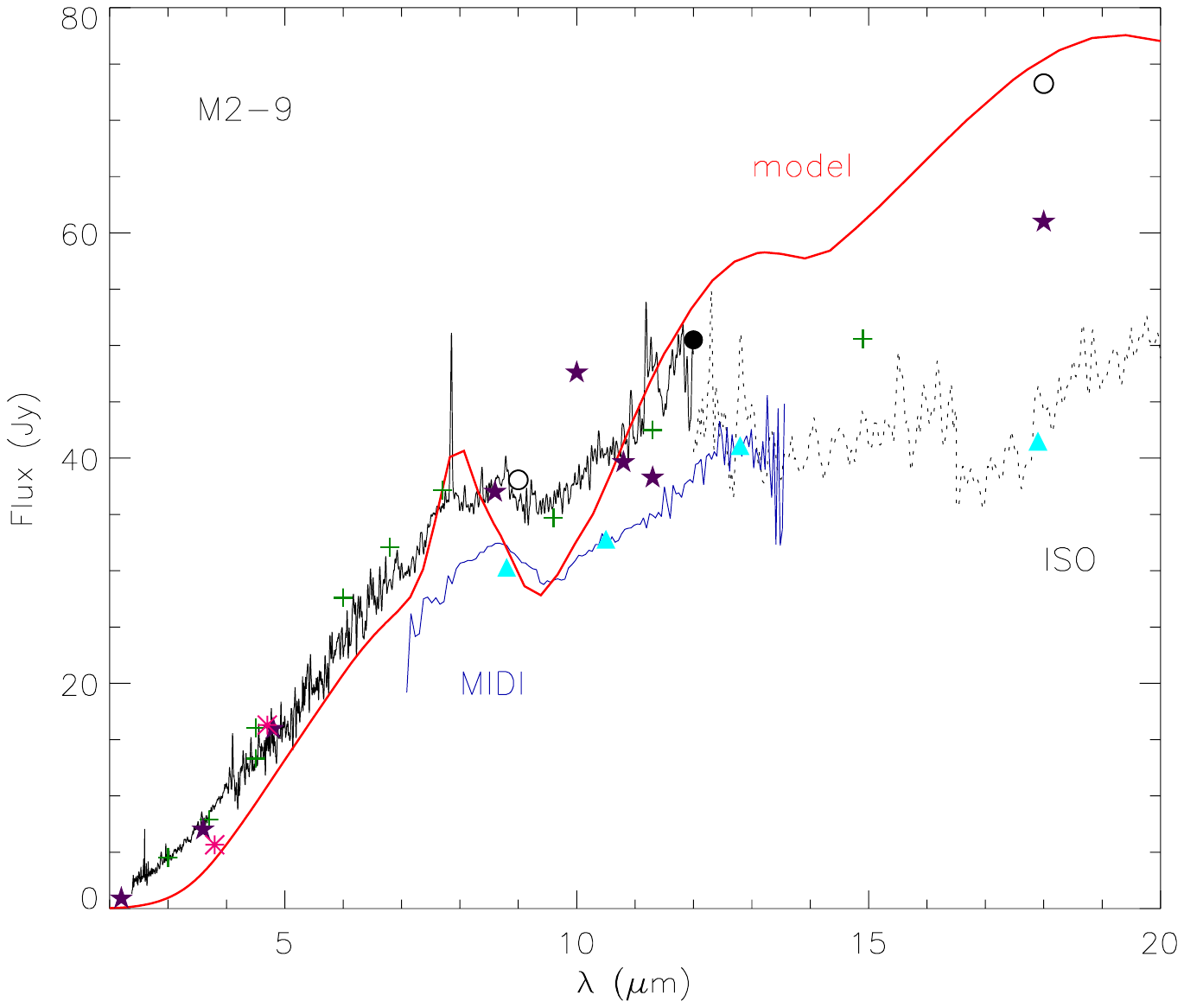}
\caption[]{At 1.2kpc: MIDI spectrum (blue) compared with the ISO spectrum (black -- solid: 14\arcsec$\times$20\arcsec aperture, dashes: 14\arcsec$\times$27\arcsec aperture), ISOCAM data (green crosses), AKARI data (open circles), the TIMMI2 photometric measurements of \citet{2005AJ....129..969S} with an aperture of 4\arcsec (cyan triangles), IRAS measurements (circles) and \citet{1974ApJ...193..401C} photometry (purple stars, 11\arcsec~beam). The NACO L$'$ and M$'$ band are also included (magenta stars, aperture 0.4\arcsec). The red line is the best model with the full aperture ($\sim$1\arcsec). All spectra have been dereddened using the~\citet{1979ARA&A..17...73S} law. \label{fig:SED}     }
\end{figure*}

\begin{figure*}
 \centering
\includegraphics[width=0.3\textwidth]{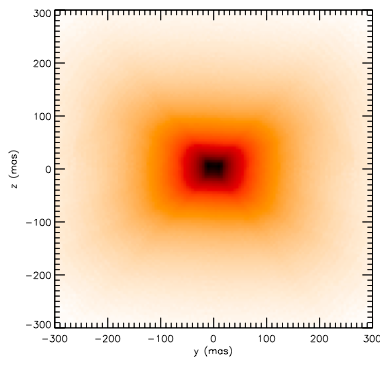}
\includegraphics[width=0.3\textwidth]{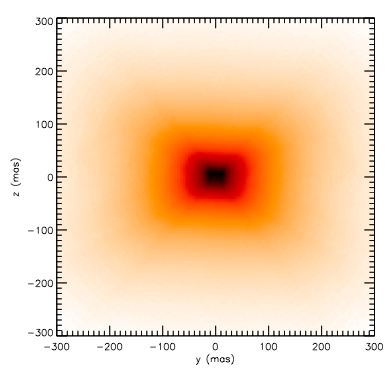}
\includegraphics[width=0.3\textwidth]{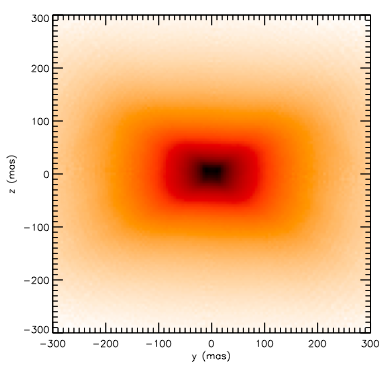}
\includegraphics[width=0.3\textwidth]{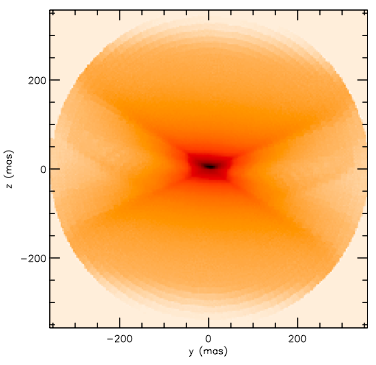}
\includegraphics[width=0.3\textwidth]{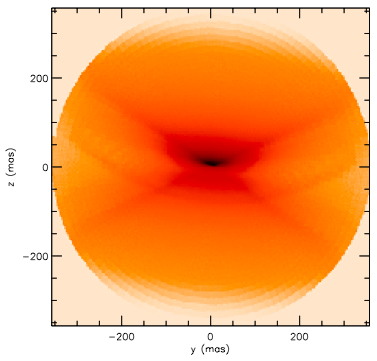}
\includegraphics[width=0.3\textwidth]{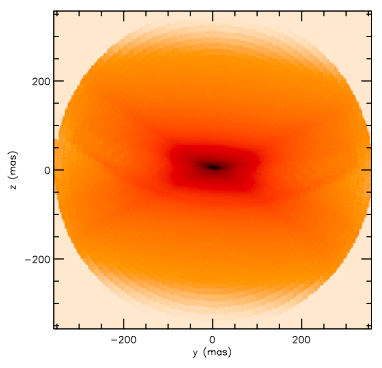}
 \caption[]{Flux distribution of the model at 8, 10 and 13 $\rm \mu m$. Images on the top belong to M2-9 and the ones on the bottom to Mz3. A small amount of light emerges at 8 $\rm \mu m$ while the flux comes from inner disk parts at 13 $\rm \mu m$.
\label{fig:maps}}
\end{figure*}

The distance of 640pc suggested by~\citet{1997A&A...319..267S} led us to many difficulties in fitting the observations. Even with scaling down the luminosity and the size of the disk or modifying the dust content, thus the optical depth, no improvement was found in fitting both the SED and the visibilities. Significant geometrical structures should have been detected in the dispersed visibilities for such a nearby source, yet the observed dispersed visibilities are very smooth. None of our models could fit the distance and luminosity assumed so far. Our best fits were found for a larger distance, namely $\sim$1200pc (Table~\ref{tab:param}, Figures~\ref{fig:SED},~\ref{fig:VIS} and \ref{fig:maps}). Recent calculations\footnote{Revision of the expansion data.} put M2-9 at a distance of 1300~$\pm$~200pc (R. Corradi, private communication).

The mass of dust, density and geometric parameters used in this model, represented the best fit for the opaque structure detected by MIDI in our line-of-sight (74\degr). A similar inclination of 73\degr~was also found independently by~\citet{2000A&A...354..674S} for the polar lobes and by~\citet{1997A&A...324..624Z} for the CO torus.

To fit the near-infrared flux, we have selected a cooler star (15,000~K) as the illuminating source than what was proposed by~\citet{2005AJ....129..969S} (30,000~K). From Fig.~\ref{fig:SED} one can notice that the shape of both MIDI and ISO spectra is the same; the only difference is the flux level. We preferred to fit the ISO spectrum, because our model has a larger aperture than that of MIDI. Our best fit for the MIDI visibilities was found by optimising the model's total $\rm \chi^{2}$, taking into account the restricted field-of-view of MIDI. The current fit has a total $\rm \chi^{2}=3.17$ (Fig.~\ref{fig:VIS}).

\begin{table}
\centering
        \caption{Parameters of the best fitted model for M2-9 and Mz3~\citep{2007A&A...473L..29C}.}\label{tab:param}
                \begin{tabular}{c|c|c}
                \hline\hline
                Parameters	& M2-9 & Mz3 \\ \hline
                T$_{\rm eff}$ (K) & 15,000 & 35,000\\
                Luminosity (L$_{\sun}$) & 2,500 & 10,000\\
                Distance (kpc)& $\sim$1.2 &$\sim$1.4 \\                \hline
                Inclination (\degr) & 74 $\pm$ 1 & 74 $\pm$ 3 \\
                best P.A. (\degr)& --2 $\pm$ 2  & 5 $\pm$ 5 \\
                Inner radius (AU) &  15 $\pm$ 1 & 9 $\pm$ 1\\
                Outer radius (AU) &  900 & 500 \\
                $\alpha$ & 2.0 $\pm$ 0.1 & 2.4 $\pm$ 0.1 \\
                $\beta$ & 0.9 $\pm$ 0.1 & 1.02 $\pm$ 0.02 \\
                $h_{\rm 100AU}$ (AU)& 36 $\pm$ 2 & 17 $\pm$ 2\\
                $\rho_{\rm grain}$ ($\rm g\,cm^{-3}$)&  2.7 & 2.7\\
                Dust mass (M$_{\sun}$) &  $1.5\times 10^{-5}$ $\pm$ $5\times 10^{-6}$ & $9\times 10^{-6}$ $\pm$ $2\times 10^{-6}$ \\ \hline
                \end{tabular}
\end{table}

We preferred to fit the visibilities rather than achieve a better fit for the SED. We could not fit our data by decreasing the outer radius, since these modifications increased the silicate absorption feature and lowered the visibilities, suggesting either a denser or a larger, resolved structure (without altering the shape of the spectrum that was due to colder dust). Expanding the outer radius imposed a new problem: more cold dust must be deposited in the disk, thus the predicted flux of the model exceeds actual measured levels at wavelengths longer than 12$\rm \mu m$. For our best fit, the disk's optical depth is in the N band is $\tau_{10\rm \mu m}=4.6$. The fit is sensitive to parameters $\alpha$, $\beta$ and $h_{0}$, as small changes in those values instantly altered the disk's opacity. The model in use does not include a puffed-up inner rim.

A single illuminating source is assumed for the dusty disk; in this case the red giant. Thus, the contribution from the companion in both the heating and shaping processes is not included and cannot be reproduced. This drawback may partly explain the difficulty in fitting the spectral energy distribution. The binary interactions lead us to consider a different disk geometry.

Due to the limited number of baselines, our selection of geometric parameters was restricted. We have found an acceptable disk model to fit the observational data, which implies a general agreement in terms of emission and spatial distribution of the bulk of the dust in the core of the system. Despite this, it is also true that the source is complex, probably engulfing a binary, and that the passive disk model is probably not a good description of the reality. This source clearly requires interferometric imaging either with aperture masking techniques~\citep{2006SPIE.6272E.103T} or with MATISSE/VLTI~\citep{2008SPIE.7013E..70L}.


\section{The mass of the components and an accretion disk}

Spectrophotometry of the core has shown an intrinsic amount of (high-excitation) iron emission lines~\citep[\ion{[Fe}{II]} and \ion{[Fe}{III]},][]{1972ApJ...174..583A}. This gives an electron temperature for the core $\sim$10,000~K and an electron density of $10^{7}\,\rm cm^{-3}$~\citep[similar to Mz3,][]{2003MNRAS.342..383S}. As it is mentioned in that work, at such high core densities collisional de-excitation prevails and weak forbidden nitrogen lines in the core are not a surprise. Iron was found in the northern lobe as well (Fig.~\ref{fig:spitzer}).~\citet{1994ApJ...437..281H} have also detected iron lines that have a high ionization potential, evidencing the presence of a particle beam or supersonic shock interface inside M2-9. H$\alpha$ emission from the core is optically thick with broad wings due to scattering~\citep{1989AJ.....97..476B,0067-0049-147-1-97}. Polarisation measurements in the optical revealed an elliptical polarisation in the core of M2-9, which was consistent in every observed waveband (namely B, R and I).~\citet{1984A&A...134..333A} explain this as scattered continuum from the central ionizing source by an extended ionized dusty torus around the core (diameter~$\lesssim4$\arcsec). All this indicates that the illuminating source of M2-9 must be a hot and compact star.

\begin{table}
\centering
\caption{Accretion luminosities for a range of WD masses and different companion mass outflows.\label{tab:lumos}}
	\begin{tabular}{c|ccccc} \hline \hline
	outflow & & & L$_{\rm acc}$~(L$_{\sun}$) & & \\
	($\rm{km\,s^{-1}}$) & 0.6~M$_{\sun}$  &  0.8~M$_{\sun}$ &  1.0~M$_{\sun}$ &  1.2~M$_{\sun}$ &  1.4~M$_{\sun}$ \\
	\hline
	5       & 1,144  &   2,760 & 5,460 & 8,734 & 14,740 \\
	10      & 73 &   174   & 338  & 566  & 914  \\ \hline
	\end{tabular}
\end{table}

A lifetime of a planetary nebula is $\sim$10,000 years and that of a white dwarf billions of years. Thus, material ejected by what is now the secondary, a white dwarf, is already dissolved in the interstellar medium. We expect that the material that composes M2-9 at the moment, originates from what is now the primary, an evolved giant, either an AGB or a post-AGB star. 

According to \citet{2008MNRAS.391.1063A} shaping of the ejecta of a tight-waist PN is possible by an accretion disk around the secondary. An accretion disk would be bright enough and illuminate part of the disk that has been detected by MIDI. We believe that the disk found in the core of M2-9, engulfs an evolved star (AGB or post-AGB) and a WD, namely the primary and secondary components of the binary. The first one is the main heating source for the dust at $\sim$2,500~L$_{\sun}$~(Table~\ref{tab:param}), while the latter obtains its luminosity from an  accretion disk, although it is the main ionizing source for the nebula (lighthouse beam). 

The most probable scenario to explain the observables of~\citet{1997A&A...319..267S} is that the mass-losing star transfers material onto the secondary at a rate of $\dot{M}_{*}=10^{-6}$~M$_{\sun}$yr$^{-1}$ at supersonic speeds (5--10$\rm{km\,s^{-1}}$) feeding the accretion disk of a white dwarf. A luminosity of $\sim$100~L$_{\sun}$ was estimated for the ionizing source from radio observations~\citep{1982MNRAS.198..321P}. The following equation by~\citet{2004MNRAS.350.1366S} was used in order to calculate accretion luminosities for the white dwarf:
\begin{eqnarray}
	\dot{M}_{\rm acc}=3\times10^{-7}\left(\frac{M_{\rm WD}}{0.6{\rm M_{\sun}}}\right)^{2}\left(\frac{v_{s}}{10\rm{km\,s^{-1}}}\right)^{-4} \nonumber \\
	\left(\frac{a}{100\rm{AU}}\right)^{-2}\left(\frac{\dot{M}_{*}}{10^{-4}\rm M_{\sun}yr^{-1}}\right)
\end{eqnarray}
where $\dot{M}_{\rm acc}$ is the mass accretion rate, $\dot{M}_{*}$ is the mass-loss rate of the primary, $M_{\rm WD}$ is the mass of the accreting white dwarf, $a$ is the binary separation and $v_{s}$ is the wind speed.

Table~\ref{tab:lumos} displays a range of accretion luminosities ($L_{\rm acc}$=G$\dot{M}_{\rm acc}M_{\rm WD}/R_{\rm WD}$) for different secondary masses for the above-mentioned accretion rates and velocities, assuming a binary separation of 30~AU and a WD radius of 6,000~km. Our findings confirm a WD mass of 0.6--0.8~M$_{\sun}$, previously suggested by~\citet{2001ApJ...552..685L} and in accordance to~\citet{2007A&A...467L..29G}.  \par We can see that the accretion luminosities for a wind velocity $\sim$10$\rm km\,s^{-1}$ aren't high enough to evaporate dust up to 15~AU (inner rim). The cool primary's luminosity (2,500~L$_{\sun}$) is sufficient though.


\section{Where is the companion relative to the dust source?}

It has long been believed that M2-9 engulfs a binary system. The light-house effect is an indirect, but rather convincing evidence of binarity. The assumption of circularity is based on the fact that no evident acceleration nor deceleration of the lighthouse's angular velocity  has been observed to date (R. Corradi, private communication)\footnote{Bearing in mind though, that intensive monitoring of the nebula occurred only in the last 20 years~\citep{2000AJ....119.1339D}}.~\citet{2001ApJ...552..685L} have used a typical mass for a white dwarf of 0.6~M$_{\sun}$ and 0.8~M$_{\sun}$ for a post-AGB star, leading to an orbital separation of 27~AU, assuming a circular orbit. 

The VLTI observations of a compact dusty structure impose some constraints, which are not trivial for the interpretation of this system. However, these interferometric observations are also limited in the sense that they do not provide any closure phase, nor absolute astrometry as in radio interferometry. It is therefore not possible to locate precisely the position of the compact dusty structure.
\subsection{The inner dusty disk.}

The size of the dusty structure detected by MIDI by the baseline closer to the equatorial plane (M2-9\_1; Table~\ref{table-log}) is approximately 33~mas (Fig.~\ref{fig:gauss}). Extrapolating to a projected baseline at 90\degr~(equatorial plane) we would expect a minimum size of 41~mas. From that we presume that the central cavity must be smaller than this size. Our modelling has showed that a diameter of approximately 30$\pm$2 AU (or $\sim$25~mas) fits the inner portion of the disk (Table~\ref{tab:param}). We did not find a better fit to the MIDI data than the one presented here by increasing the disk's inner rim.

We have used a relation for dust sublimation radius ($R_{\rm sub}$ in AU) by~\citet{2001Natur.409.1012T}:
\begin{equation}
	R_{\rm sub}=1.1\left(\frac{T_{\rm sub}}{1500\rm K}\right)^{-2}\sqrt{\frac{L}{1000\rm L_{\sun}}}
\end{equation}
where $L$ is the stellar luminosity and $T_{\rm sub}$ is the dust sublimation temperature. For a star of 2500~L$_{\sun}$ (Table~\ref{tab:sublim}) it is suggested that the inner rim of the disk must reside at a distance larger than 4~AU but within the limits defined by the MIDI observations. An 100~L$_{\sun}$ heating source, such as the accretion disk in the case of M2-9, does not fit the observational data; neither did the 553~L$_{\sun}$ suggested by~\citet{1997A&A...319..267S}.

\begin{table}
        \caption{Dust sublimation radii in~AU for two stellar luminosities \label{tab:sublim}  }
        \centering
                \begin{tabular}{lcc}
                \hline \hline
                T$_{\rm sub}$ [K]	& \multicolumn{2}{c}{R$_{\rm sub}$ [AU]} \\
                 & 2500~L$_{\sun}$ & 100~L$_{\sun}$ \\ \hline
                2000 	& 0.97	&	0.19	\\
                1500 	& 1.74	&	0.35	\\
                1000 	& 3.90	&	0.78	\\
		\hline
                \end{tabular}
\end{table}

We estimated orbital periods for the binary by using the observations of 18 years from \citet{2000AJ....119.1339D}. From those, it can be seen that the lighthouse beam covered less than 20\% of the perimeter of the lobes within that time scale. We found that the period ranges from 90 to 120 years.

A binary system should reside within the structure detected by MIDI. By applying simple Keplerian physics (assuming circular orbits), we have used several mass ratios\footnote{Within the Chandrasekhar limit for the WD} for the binary components to estimate the corresponding orbital diameters, for periods of 90 and 120 years, in physical units and converted them to angular sizes for a distance of 1.25~kpc (Table~\ref{tab:binsep}). 

As seen in Table~\ref{tab:binsep} orbital diameters that fit within the extrapolated angular size of 41~mas range from 42--52 AU. This is larger than our best fit model, where the size of the disk's inner cavity is 30~AU. We should draw attention though to the fact that the disk's inner rim is centered at the heating source, which in turn is not positioned on the binary's centre-of-mass. The model in use cannot reproduce a partially illuminated disk or a disk with an off-centre illuminating source. Thus, we believe that the companion is truncating the disk, since its heating capabilities are minimal.

We can only surmise that high-angular resolution radio observations could establish the astrometric position of the companion (WD/accretion disk).

\begin{table}
\centering
\caption{Binary orbital diameters in AU (also converted in mas for a distance\,=\,1.25~kpc), at two different periods, 120 and 90 years}\label{tab:binsep}
        \begin{tabular}{cc|cc|cc}\hline \hline
        M$_{1}$ (M$_{\sun}$)    &   M$_{2}$ (M$_{\sun}$)     & \multicolumn{2}{c|}{P = 120 yrs}   & \multicolumn{2}{c}{P = 90 yrs}\\  
        &  &  d~[AU]   & d~[mas] &   d~[AU] & d~[mas] \\
        \hline
        0.6  &        0.6        &  52     &   41.6  	& 42	&	33.6\\
        0.8  &        1.0        &  60     &   48.0	& 50	&	40.0\\
        0.6  &        1.4        &  62     &   49.6	& 50	&	40.0\\
        0.8  &        1.4        &  64     &   51.2	& 52	&	41.6\\  
        1.0  &        1.2        &  64     &   51.2	& 52	&	41.6\\   
        0.6  &        1.0        &  57     &   45.7	& 47	&	37.7\\
        0.6  &        1.2        &  60     &   48.0	& 49	&	39.2\\
        1.0  &        1.0        &  62     &   49.2	& 51	&	40.6\\
         \hline
        \end{tabular}
\end{table}

\subsection{The exterior torus}
The obtained NACO images are fully dominated by the central source and lack dynamical range in its close vicinity (Fig.~\ref{fig:hst-naco}). At $\sim$3\arcsec~from the core~\citet{1997A&A...324..624Z} found that the CO emission is coming from a large ring, whose center seemed offset by 0.5$\pm$0.3\arcsec~from the compact radio source~\citep[$<0.1$\arcsec~and unresolved at 1.3~cm,][]{1983IAUS..103...69B}. More recent Plateau de Bure data (Castro-Carrizo et al., in preparation) with 0.3\arcsec~angular resolution confirm the presence of an offset, and detect additional CO emission coming from regions at a distance of $\sim$1\arcsec~from the center.

Molecular hydrogen spectroscopy has yielded two different velocity components at distances $\lesssim$0.5\arcsec~latitudinally from the central illuminating source: 10.9$\rm km\,s^{-1}$ (blue-shifted) and 123.7$\rm km\,s^{-1}$ (red-shifted). This points to a H$_{2}$ disk-like structure~\citep{2005AJ....130..853S}.

Spiral structures (as expected in a symbiotic type system) can be formed by binary interactions during the ejection of the primary's CO envelope~\citep{2008ApJ...675L.101E}, yet this is not clear at this point for the case of M2-9. If there are spirals, they are seen at a high inclination (pole-on). Our source is edge-on. The differential phases of the disk detected by VLTI are small, lying within a range of $\pm$10\degr, i.e. well below some strong signature from dusty rings reported in the literature~\citep{2007A&A...474L..45D, 2008A&A...490..173O}.


\section{Comparing disks: M2-9 and Menzel 3}

These two bipolar nebulae are spectroscopic twins and both exhibit complex gaseous shell structures. Both are extended at similar sizes. Their main parts, including the lobes, are 43\arcsec~and 38\arcsec~at length and $\sim$15\arcsec~at width, for M2-9 and Mz3 respectively. The ansae of M2-9 are 115\arcsec~apart, while the farther regions of Mz3 are at about 130\arcsec.

Distances to both objects are not well determined and the estimates given stand for the best fits to the observational data: M2-9 at 1.2kpc, Mz3 at 1.4kpc. Indirect evidence for binarity in the case of M2-9 are the lighthouse knots in the nebula and the fast outflows observed in H$\alpha$ (as recently studied in the case of the Red Rectangle, \citet{2009ApJ...693.1946W}), and for Mz3, the X-ray jet~\citep{2003ApJ...591L..37K} and also fast outflow seen in H$\alpha$. 

\begin{figure}[!h]
 \centering
    \includegraphics[width=0.8\textwidth, angle=90]{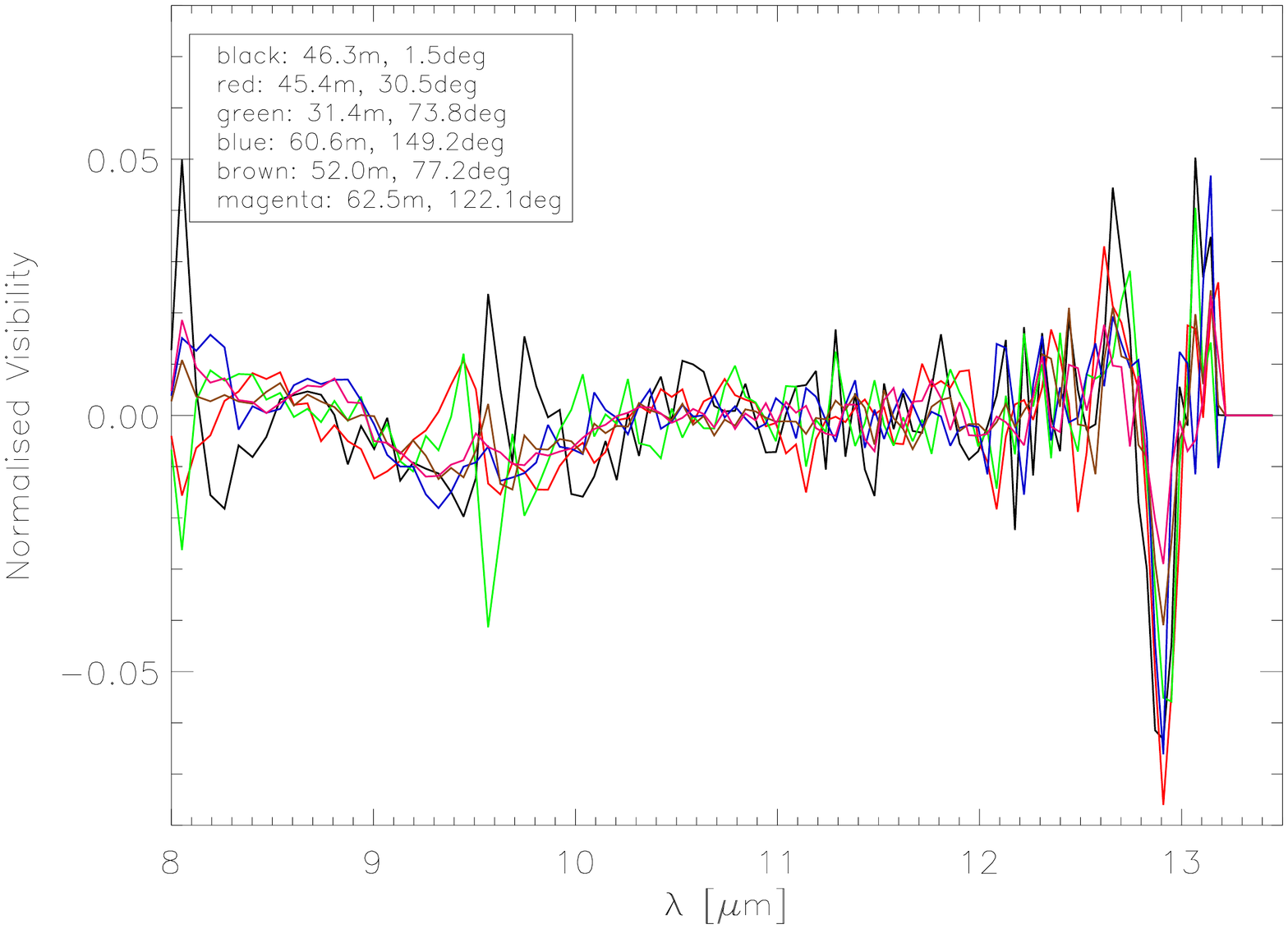}
    \caption[]{Normalized visibilities vs. wavelength for Mz3~\citep{2007A&A...473L..29C}. The prominent feature belongs to the bright 12.8$\rm \mu m$ [NeII] nebular emission line. The noise increase at 9.7$\rm \mu m$ corresponds to the narrow ozone feature. Observed baselines are indicated within the legend. \label{fig:mz3freq} }
\end{figure}

Whereas Mz3 was unresolved in all infrared filters, some significant extensions of the nebular core at larger diameters in both equatorial and polar directions were found for M2-9, as already suspected in some other observations. Our observations in the mid-infrared have revealed that they both have very compact dusty disks in their cores.

The disks have been resolved with high-angular resolution techniques and their composition is similar (amorphous silicates). In addition, the disks are in the equatorial planes of the planetary nebulae (that is, perpendicular to the bipolar ejecta) and at similar inclination from the rotation axis ($\sim$74\degr).

The disk inside M2-9 is larger in terms of size (Fig.~\ref{fig:maps}) and hot dust mass compared to Mz3: $1.5\times 10^{-5}$~M$_{\sun}$ and $9\times 10^{-6}$~M$_{\sun}$, respectively. In both cases, the total amount of material is 0.5--1~M$_{\sun}$, which points to progenitors of several Solar masses. This is in line with the high nitrogen content found by~\citet{2003MNRAS.342..383S} for Mz3. In addition, the disk of M2-9 exhibits some degree of crystallinity (Fig.~\ref{fig:forst}), which may indicate that it is older, or that the physical conditions for annealing prevail. There are no such features in Mz3 (Fig.~\ref{fig:mz3freq}).

In the work of~\citet{2003MNRAS.342..383S}  it is shown that the optical to infrared SED of Mz3 required two components; a hot star and a cool star. M2-9 has a very similar SED in the optical to that of Mz3, thus it is probable that it has similar components. \par
Assuming a circular orbit might be an oversimplification. Binaries containing a WD and another evolved star, may have eccentric orbits. When main sequence stars will ascend to the red giant branch and then to the AGB, it should be expected that despite strong mass-loss the eccentricity will persist, leading to variations in the building up of an accretion disk and the generation of jets. This may lead to short and strong events of large mass ejection, related to the building of large scale structures seen in the nebula, and in the equatorial plane as well.

\section{Conclusions}

We present the discovery of a complex dusty disk inside the elongated bipolar planetary nebula M2-9, with the use of infrared interferometry, for which we have derived geometrical constraints and chemical composition. Our findings show that the disk is much smaller than the extended dusty structure around the nebula's core with an inner rim of 15AU at 1.2kpc and that it is primarily composed of silicates. This suggests that the dusty material is derived from the AGB envelope of the primary and has survived its evolution onwards. The disk is also aligned with the equatorial axis of the nebula. It is probable that the dust is heated by the primary, evolving star (0.6--1.4M$_{\sun}$), but it is also partially heated and truncated by the accretion disk of the secondary, a white dwarf (0.6--1M$_{\sun}$).\par
There are similarities between the disk found in M2-9 and that of its spectroscopic twin, Mz3. Although the sample is small, many bipolar planetary nebulae and proto-planetary nebulae (e.g. M2-9, Mz3, OH231.8+4.2, He2-113, CPD-56\degr8032) have dusty structures (disks/tori/spirals) around their cores. Some of them are known to engulf binary systems. Binarity in the cores and dusty structures around them may explain the shaping of the ejecta into bipolar shapes. Thus, an extended study of those bipolar proto-planetary and planetary nebulae would be advised to establish whether this is a global tendency.\par
M2-9 is yet another example of an object with silicate chemistry within the disk, suggesting that the material was ejected during the oxygen-rich phase of the primary star. Nevertheless, there are PAHs in the lobes of the nebula, at a much lower percentage, most probably formed after the dissociation of carbon monoxide.


\begin{acknowledgements}
      F.L. would like to thank: the referee for his constructive comments, the Fizeau Exchange Visitors Program of the European Interferometry Initiative for supporting part of this work, the ISO helpdesk for providing the CIA software and A. Duarte-Cabral and C. Gielen for their help.
\end{acknowledgements}

\bibliographystyle{aa} 
\bibliography{references}  

\end{document}